\documentstyle[12pt]{article}
\begin{document}

\title{Recurrent procedure for the determination of
the Free Energy $\epsilon^{2}$-expansion in the Topological
String Theory.}

\author{B.A.Dubrovin \,\, , \,\, A.Ya.Maltsev\thanks{also
L.D.Landau Institute for Theoretical Physics, Russian Acad. Sci.,
maltsev@itp.ac.ru} }

\date{SISSA-ISAS, Via Beirut 2-4 - 34014 Trieste - Italy,
\,\,\, dubrovin@sissa.it \,\, , \,\, maltsev@sissa.it }

\maketitle

\begin{abstract}
We present here the iteration procedure for the determination
of free energy $\epsilon^{2}$-expansion using the theory of
KdV - type equations. In our approach we use the conservation
laws for KdV - type equations depending explicitly on times
$t_{1}, t_{2}, \dots$ to find the $\epsilon^{2}$-expansion
of $u(x,t_{1},t_{2},\dots)$ after the infinite number of
shifts of $u(x,0,0,\dots) \equiv x$ along $t_{1}, t_{2}, \dots$
in recurrent form. The formulas for the free energy expansion
are just obtained then as a result of quite simple integration
procedure applied to $u_{n}(x)$.
\end{abstract}

This work is devoted mainly to the calculation of 
low-dispersion expansions of the solutions of KdV
type equations and their using for calculation of
Weil-Petersson volumes of moduli spaces. More precisely
we obtain a recurrent procedure for the formulas 
presented in ~\cite{EYY1} for such expansions and
will refer here to the papers ~\cite{EYY1} and
~\cite{zog1} where the more detailed information
and references can be found. Our procedure is based
on the quasi-classical expansion for Schr\"odinger
operator and uses also the times-dependent integrals
for KdV type equations which, as far as we know,
were not mentioned in the previous papers.

We shall consider the Free Energy function of Topological
String Theory $F(x,t_{1},t_{2},\dots)$ such that its second
derivative with respect to $x$ \linebreak 
$u(x,t_{1},t_{2},\dots) = {d^2 \over dx^2} F$
satisfies at any 
$t_{1},t_{2},\dots$ the KdV hierarchy with respect to all
$t_{1},t_{2},\dots$. 

The first KdV-equation is written here in the form:

\begin{equation}
\label{kdv}
{\partial u \over \partial t_{2}} = u u_{x} + 
{ {\epsilon}^{2} \over 12} u_{xxx} 
\end{equation}

where $\epsilon$ is small parameter, so we can consider the
$\epsilon^{2}$-expansion of the solution $u(x,t_{1},t_{2},\dots)$
of KdV hierarchy and after that try to get $F(x,t_{1},t_{2},\dots)$
by the integration with the corresponding normalizing conditions.

We shall describe now the construction which permits to obtain
the $\epsilon^{2}$-expansion of $u(x,t_{1},t_{2},\dots)$
using times-dependent conservation laws \linebreak
$J(t_{1},t_{2},\dots)$ for KdV hierarchy.

It is well known that (\ref{kdv}) can be written in the form:

\begin{equation}
\label{lax}
{\partial L \over \partial t_{2}} = [A_{2},L]
\end{equation}

\begin{equation}
\label{l}
L = - {{\epsilon}^{2} \over 12} {d^{2} \over dx^{2}} - {u(x) \over 6} 
\end{equation}

\begin{equation}
\label{a}
 A_{2} = {\epsilon^{2} \over 3} {d^{3} \over dx^{3}} +
{1 \over 2} \left(u {d \over dx} + {d \over dx} u \right) 
\end{equation}

Equation $L\psi = {k^{2} \over 12}$ after the substitution:

\begin{equation}
\label{hi}
i \chi (x,k,\epsilon) = {1 \over \epsilon} {d \ln \psi \over dx}
\end{equation}

takes the form:

\begin{equation}
\label{hieq}
- i \epsilon {d \over dx} \chi + {\chi}^{2} = k^{2} + 2 u(x) 
\end{equation}

and using this form we can obtain the densities of conservation 
laws for (\ref{kdv}) from the formal expansion:

\begin{equation}
\label{hiexp}
\chi (x,k,\epsilon) \sim k + \sum_{n=1}^{\infty}
{\chi_{n}(x,\epsilon) \over (2k)^{n}} 
\end{equation}

when $k \rightarrow \infty$. Here $\chi_{2n}(x,\epsilon)$ are
full derivatives 
$\chi_{2n}(x,\epsilon) = \partial_{x} Q_{2n}(x,\epsilon)$, and 
$\chi_{2n+1}(x,\epsilon)$ - are densities of conservation laws,
which are polynomial expressions of 
$u, \epsilon u_{x}, {\epsilon}^{2} u_{xx}, \dots $

$$ I_{n} = \int \chi_{2n+1}(x,\epsilon) dx  $$

We can write the higher KdV-type equations, consisting with 
(\ref{kdv}),  in the Hamiltonian form:

\begin{equation}
\label{vyssh}
 {\partial u \over \partial t_{k}} = \partial_{x} 
{\delta \over \delta u(x) } \int \chi_{2n+1}(x,\epsilon) dx =
\partial_{x} {\delta \over \delta u(x)} {1 \over \pi i}
\oint (2k)^{2n} \left[ \int \chi (x,k,\epsilon) dx \right] dk 
\end{equation}

Theorem 1.

Let us consider equations (\ref{vyssh}) on the functional space of
rapidly decreasing functions: 
$u(x) \rightarrow 0 , |x| \rightarrow \infty$.
Then the functional:

\begin{equation}
\label{int}
J_{n}(t_{n}) = \int x u(x) dx - 4(2n - 1) t_{n}
\int \chi_{2n-1}(x,\epsilon) dx
\end{equation}

is the conservation law for n-th KdV-equation (\ref{vyssh}),
depending on the time $t_{n}$.\footnote{Firstly the functionals of
this type were considered in ~\cite{dubr1} , where they
were restricted on the slowly modulated m-phase algebro-geometric
solutions of KdV. Here we derive the analogous formulas for any solution
of KdV-type equations.} 
                          
Proof.

Since $\int \chi_{2n-1}(x,\epsilon) dx$ is the conservation law
for any of equations (\ref{vyssh}), we have:

$$ {d \over dt_{n}} J_{n} = \int x \partial_{x} 
{\delta \over \delta u(x)} \int \chi_{2n+1}(y,\epsilon) dy dx -
4(2n-1) \int \chi_{2n-1}(x,\epsilon) dx =$$
$$= - \int dx  {\delta \over \delta u(x)} I_{n} -
4(2n-1) I_{n-1}  $$

But as can be extracted from (\ref{hieq}) and (\ref{hiexp}):
\footnote{For the identities of this type see also
~\cite{gd}.}

$$ \int   {\delta \over \delta u(x)} I_{n} dx  \equiv 
\int {\delta \over \delta u(x)} {1 \over \pi i} (2k)^{2n}
\left[ \int \chi (y,k,\epsilon) dy \right] dk dx =$$ 
\begin{equation}
\label{sootnint}
= {1 \over \pi i} \oint (2k)^{2n} \left[ 2 {d \over d(k^2)}
 \int \chi (y,k,\epsilon) dy \right] dk = - 4(2n-1) I_{n-1} 
\end{equation}
so we have ${d \over dt_{n}} J_{n} = 0$.

Theorem is proved.

Let us now consider the variational derivative of $J_{n}$ with 
respect to $u(x)$ corresponding to rapidly decreasing variations 
$\delta u(x)$, that is the variational derivative of 
Euler-Lagrange type:

\begin{equation}
\label{vder}
\Omega_{n}(x,\epsilon) = {\delta J_{n} \over \delta u(x)} 
\equiv {\partial P_{n} \over \partial u}(x) - 
{\partial \over \partial x} {\partial P_{n} \over \partial u_{x}}(x) +
{\partial^{2} \over \partial x^{2}} 
{\partial P_{n} \over \partial u_{xx}}(x) - \dots
\end{equation}

where 
$J_{n} = \int P_{n}(u,\epsilon u_{x}, {\epsilon}^{2} u_{xx},\dots) dx$ .

Theorem 2.

If we consider the n-th equation of (\ref{vyssh}) then the evolution of
$\Omega_{n}(x,\epsilon)$ satisfies the linear equation:

$${d \over dt_n} \Omega_{n}(x,\epsilon) = 
D_{n}^{1}(\epsilon,u, u_{x},\dots) {d \over dx} \Omega_{n} +
\dots +
D_{n}^{2n-1}(\epsilon,u, u_{x},\dots) 
{d^{2n-1} \over dx^{2n-1}} \Omega_{n} =$$
\begin{equation}
\label{omegaev}
= \sum_{s=1}^{2n-1} 
D_{n}^{s}(\epsilon,u, u_{x},\dots,u_{(2n-s-1)x}) 
{d^{s} \over dx^{s}} \Omega_{n}
\end{equation}
where $D_{n}^{s}(\epsilon,u, u_{x},\dots)$ are some polynomials of
$u, u_{x}, u_{xx}, \dots$, ${d \over dt_{n}}$ is full derivative with
respect to $t_{n}$, $u_{nx} \equiv {d^{n} \over
dx^{n} } u(x)$.

Proof.

Let us consider any of equations (\ref{vyssh}) as the flux on the space
$(u(x))$ corresponding to vector field:

$$\xi (x) = \partial_{x} {\delta \over \delta u(x)} 
\int \chi_{2n+1} (x,\epsilon) dx  $$

Since $J_{n}$ is invariant function, then $\Omega (x,\epsilon)$
is the invariant 1-form on the space $(u(x))$. So its full
Lie-derivative with respect to $\xi(x)$ plus partial derivative
with respect to $t_{n}$ must be zero, that is:

$${\partial \over \partial t_n} \Omega_{n}(x,\epsilon) + 
\left( L_{\mbox{$\boldmath \xi$}} \Omega_{n} \right) (x) = 0 $$
where
$$\left( L_{\mbox{$\boldmath \xi$}} \Omega_{n} \right) (x) =
\int \xi(y) {\delta \over \delta u(y)} \Omega_{n}(x) dy +
\int \Omega_{n}(y) {\delta \over \delta u(x)} \xi(y) dy $$

and

$$ {\partial \over \partial t_n} \Omega_{n}(x,\epsilon) +
\int \xi(y) {\delta \over \delta u(y)} \Omega_{n}(x) dy \equiv
{d \Omega_{n}(x) \over dt_{n}} $$

The expression $\int \Omega_{n}(y) {\delta \over \delta u(x)} \xi(y) dy$
is the action of the linear differential operator of type (\ref{omegaev}) 
on $\Omega_{n}(x)$.

Theorem is proved.

Corollary.

If  $\Omega_{n}(x) = 0$ at $t_{n} = 0$ (and u(x) is rapidly
decreasing), then  $\Omega_{n}(x) \equiv 0$ at any $t_{n}$.

There can be easily formulated the generalizations of Theorems
1 and 2 if we consider the common solution 
$u(\epsilon, x, t_{1}, t_{2},\dots , t_{n}, C_{1}, C_{2},\dots , C_{n})$
of the system of equations:

\begin{equation}
\label{vsevyssh}
{\partial u \over \partial t_{n}} = C_{n} \partial_{x} 
{\delta \over \delta u(x)} \int \chi_{2n+1} (x,\epsilon) dx,
\,\,\, n = 1, \dots , N.
\end{equation}

Then:

Theorem $1^{\prime}$.

The functional:

\begin{equation}
\label{obint}
J (t_{1},\dots,t_{N},C_{1},\dots,C_{N}) = \int x u(x) dx -
\sum _{s=1}^{N} 4 (2s-1) C_{s} t_{s} \int \chi_{2s-1} (x,\epsilon) dx
\end{equation}

- is the conservation law for all fluxes (\ref{vsevyssh}) for
$1 \leq n \leq N$, that is:

$$ {d \over dt_{n}} J (t_{1},\dots,t_{n},C_{1},\dots,C_{N})
\equiv 0, \,\,\,\, 1 \leq n \leq N .$$

The proof is evident since any of 
$t_{s} \int \chi_{2s-1} (x,\epsilon)dx$,
$s \neq n$ at fixed $t_{s}$ is the conservation law for n-th KdV-equation
and all KdV-equations commute with each other.

Theorem $2^{\prime}$.

The coefficients $\Omega (x)$ of 1-form 
$\mbox{$\boldmath \Omega$}$:  $\Omega (x) = {\delta J \over \delta u(x)}$
(Euler-Lagrange derivative) satisfy the system of equations:

$${d \over dt_{n}} \Omega (x,t_{1},\dots,t_{n}) = 
\sum_{s=1}^{2n-1} C_{n} D_{n}^{s} (\epsilon, u, u_{x},\dots,u_{(2n-s-1)x})
{d^{s} \over dx^{s}} \Omega (x) $$
\begin{equation}
\label{formaev}
 n = 1,\dots,N ,
\end{equation}   
and if  $\Omega (x)$ is zero at $t_{1}=t_{2}= \dots =t_{N} = 0$
then it is identically zero at any  $t_{1},t_{2}, \dots, t_{N}$.

It is also evident that we can add to  $\Omega (x)$ any invariant
form of type

$$\Omega^{\prime} (x) = \sum_{s=0}^{M} d_{s} {\delta \over \delta u(x)}
\int \chi_{2s+1}(x,\epsilon) dx $$
(where the coefficients $d_{s}$ do not depend on $t_{1},t_{2},\dots$)
and Theorems 2, $2^{\prime}$ will remain valid.

We shall need later the invariant forms of type

\begin{equation}
\label{forma}
\Omega (x) = x - u(x) + \sum_{s=1}^{\infty} \beta_{s} t_{s}
{\delta \over \delta u(x)} \int \chi_{2s-1}(x,\epsilon) dx
\end{equation}

(where $u(x)$ is the variational derivative of the momentum
integral $P = {1 \over 2}\int u^{2}(x) dx$) 
for the investigation of asymptotic expansion of $u(x)$ in terms of
$\epsilon^{2}$ after the infinite number of shifts of the initial
function $u(x,0,0,\dots) = x$ along times $t_{1},t_{2},\dots$
according to KdV equations (\ref{vyssh}).

All the considerations above were for rapidly decreasing functions
$u(x)$. But as can be easily seen, the relations (\ref{omegaev}),
(\ref{formaev}) are local expressions of 
$u,u_{x},u_{xx},\dots,u_{t_{s}},u_{xt_{s}},u_{xxt_{s}},\dots$
where we consider 
$D_{n}^{s}, \Omega (x)$ (and \linebreak
${\delta \over \delta u(x)} \int \chi_{2s+1}(x,\epsilon) dx$)
just as local polynomials of $u,u_{x},\dots$ 
( for the last we use just formal Euler-Lagrange expression 
for variational derivative
in this case) and Theorems 2, $2^{\prime}$ will be valid for
1-forms (\ref{vder}) and (\ref{forma}) for any global in $x$
solution $u(x)$ up to the time $t_{1},t_{2},\dots,t_{N}$
where this global solution exists (so if $\Omega (x)$
is identically zero at $t_{1} = t_{2} = \dots = 0$ it will be 
identically zero in all region where we have a global solution
$u(x)$.)

\vspace{1cm}

Now we shall consider the following construction:

\vspace{1cm}

It is well known that all KdV-type equations (\ref{vyssh}) in our 
case can be written in the form:

\begin{equation}
\label{kdvtype}
{\partial u \over \partial t_{n}} = {1 \over \alpha_{n}}
u^{n-1} u_{x} + \epsilon^{2} (\dots) + \epsilon^{4} (\dots) +
\dots
\end{equation}

Let denote $K_{n}^{m}(u,u_{x},u_{xx},\dots)$ - the corresponding term
in n-th KdV equation which has the multiplier $\epsilon^{2m}$, so we 
have

\begin{equation}
\label{obshur}
{\partial u \over \partial t_{n}} = {1 \over \alpha_{n}}
u^{n-1} u_{x} + \sum_{m=1}^{n-1} \epsilon^{2m}
K_{n}^{m}(u,u_{x},u_{xx},\dots) 
\end{equation}

Let we are given the function $\Phi (u)$ which is a convergent 
everywhere (in u-plane) series:

\begin{equation}
\label{ffunction}
\Phi (u) = \sum_{n=0}^{\infty} \gamma_{n} u^{n} 
\end{equation}

We can consider the common solution of all KdV-type equations
(\ref{kdvtype}) given by the relations

$${\partial u \over \partial t_{n}} = \alpha_{n} 
\gamma_{n-1} \partial_{x}
{\delta \over \delta u(x)} \int \chi_{2n+1}(y,\epsilon) dy =$$
\begin{equation}
\label{frelations}
= \alpha_{n} \gamma_{n-1} \partial_{x} {\delta \over \delta u(x)}
{1 \over \pi i} \oint (2k)^{2n} 
\left[ \int \chi (y,k,\epsilon) dy \right] dk
\end{equation}
up to the times $t_{1},t_{2},\dots$, where this global solution
$u(x,t_{1},t_{2},\dots)$ exists and we put 
$u(x,0,0,\dots) = x$.

Theorem 3.

If $\Phi (u) = \sum_{n=0}^{\infty} \gamma_{n} u^{n}$ is a convergent
everywhere series, then the equation

\begin{equation}
\label{utauur}
{\partial u \over \partial \tau} = \sum_{n=1}^{\infty}
\alpha_{n} \gamma_{n-1} \partial_{x} {\delta \over \delta u(x)}
\int \chi_{2n+1}(x,\epsilon) dx
\end{equation}

can be represented as:

\begin{equation}
\label{epsuravn}
{\partial u \over \partial \tau} = \Phi (u) u_{x} +
\sum_{n=1}^{\infty} \epsilon^{2m} K_{m}(u,u_{x},\dots,u_{(2m+1)x})
\end{equation}

where $K_{m}(u,u_{x},\dots,u_{(2m+1)x})$ are the polynomials of
$u_{x},u_{xx},\dots$ (containing in each term (2m+1) derivatives
with respect to x) with the coefficients depending on $u$ and
being expressed in terms of $\Phi (u), \Phi^{(1)}(u), \Phi^{(2)}(u),
\dots$,
$\Phi^{(q)}(u) \equiv {d^q \over du^{q}} \Phi(u)$.

Proof.

Let us remind that the KdV-hierarchy can be extracted from the
Riccati equation (\ref{hieq}) for $\chi (x,\epsilon,k)$:

$$ -i \epsilon {d \over dx} \chi + \chi^{2} = k^{2} + 2 u(x) $$
and we use the function:

\begin{equation}
\label{hir}
\chi_{R} (x,k) \equiv k + \sum_{n=0}^{\infty} 
{\chi_{2n+1}(x,\epsilon) \over (2k)^{2n+1} }
\end{equation}  

for the generation of Hamiltonian KdV-fluxes. 

As was shown by B.A.Dubrovin (see ~\cite{dmn},
~\cite{dubr2}, ~\cite{dubr3}) the following relation
holds:

\begin{equation}
\label{dubrrel}
{\delta \over \delta u(x)} \int \chi_{R} (x,\epsilon,k) dx 
\,\,\,(\equiv {\delta \over \delta u(x)} \int \chi (x,\epsilon,k) dx) =
{\lambda \over \chi_{R}(x,\epsilon,k)}    
\end{equation}
in the class of rapidly decreasing or periodic functions $u(x)$. So
it is valid as local relation for $u,u_{x},\dots$ in any order of formal
expansion of $\chi_{R} (x,\epsilon,k)$ and 
${1 \over \chi_{R} (x,\epsilon,k)}$ in terms of $1/(2k)^{2s+1}$, where
we use just their local expressions obtained from the formal equation
(\ref{hieq}) and ${\delta \over \delta u(x)}$ is just formal
Euler-Lagrange expression for derivatives of 
$\int \chi_{R,(2n+1)} (x,\epsilon) dx$ with respect to $u(x)$:

$$ {\delta \over \delta u(x)} \int \chi_{R,(2n+1)} (x,\epsilon) dx
= {\partial \chi_{R,(2n+1)} (x,\epsilon) \over \partial u} -
{\partial \over \partial x} 
{\partial \chi_{R,(2n+1)} (x,\epsilon) \over \partial u_{x}}
+ \dots $$
and we have
$$ {\delta \over \delta u(x)} \int \chi_{R,(2n+1)} (x,\epsilon) dx =
\lambda \left[ {1 \over \chi_{R} (x,\epsilon,k)} \right]_{2n+1} $$
($[ \dots ]_{2n+1}$ means here the corresponding term in the 
expansion).

We shall not need the value of constant $\lambda$ in our case, but
what is important that $\lambda$ does not depend upon $\epsilon$.
This fact can be easily obtained from the fact that, as follows from
the formal equation (\ref{hieq}), the formulas for 
$\chi_{n}(x,\epsilon)$ in (\ref{hiexp}) as the expressions of
$u,u_{x},u_{xx},\dots$ differ from the analogous formulas at
$\epsilon = 1$ just by the multiplier $\epsilon$ in any 
differentiation with respect to x (as well known, $\epsilon$
can be removed from the initial KdV-equation (\ref{kdv})
by scaling transformation). Since the relation (\ref{dubrrel})
in any order of $k^{-1}$ is just the identical equality between
two polynomials of $u,u_{x},u_{xx},\dots$ it will remain true
if we replace any differentiation with respect to $x$ by
$\epsilon {d \over dx}$. It can be also checked by the direct
calculations similar to ~\cite{dubr2},~\cite{dubr3} 
that here $\lambda = 1$.

So we can represent the equation (\ref{utauur}) as:

$$ {\partial u \over \partial \tau} =
\lambda \sum_{n=1}^{\infty} \alpha_{n} \gamma_{n-1} 
\partial_{x} {1 \over \pi i} \oint (2k)^{2n} 
{dk \over \chi_{R} (x,\epsilon,k)} $$
and write the closed equation for
$A (x,\epsilon,k) \equiv {1 \over \chi_{R} (x,\epsilon,k)}$

\begin{equation}
\label{auravn}
{1 \over 2} \epsilon^{2} A {d^{2} \over dx^{2}} A -
{1 \over 4} \epsilon^{2} \left({d \over dx} A \right)^{2} 
= 1 - (k^{2} + 2 u(x)) A^{2}
\end{equation}
obtained from (\ref{hieq}) by substitution 
$\chi = \chi_{R} + i \chi_{Im}$, where $\chi_{R}$ is real
if $k^{2} + 2 u(x) \geq 0$ and coincides with the introduced
above, and $A = 1/\chi_{R} (x,\epsilon,k)$,
$B = \chi_{Im}(x,\epsilon,k)/\chi_{R} (x,\epsilon,k)$.

After that we obtain the following system of equations:

$${\partial u \over \partial \tau} =
\lambda \sum_{n=1}^{\infty} \alpha_{n} \gamma_{n-1}
\partial_{x} {1 \over \pi i} \oint (2k)^{2n} A (x,\epsilon,k) dk $$
\begin{equation}
\label{pervsyst}
{1 \over 2} \epsilon^{2} A {d^{2} \over dx^{2}} A -
{1 \over 4} \epsilon^{2} \left({d \over dx} A \right)^{2}
= 1 - (k^{2} + 2 u(x)) A^{2}
\end{equation}

where ${1 \over \pi i} \int (2k)^{2n} A (x,\epsilon,k) dk$ is
just the formal expression meaning that we must take the n-th
term in the formal expansion:

$$ A(x,\epsilon,k) \sim \sum_{n=0}^{\infty} 
{A_{2n+1}^{*}(x,\epsilon) \over (2k)^{2n+1}} , \,\,\,
A_{1}^{*}(x,\epsilon) \equiv 2 , $$
$k \rightarrow \infty$.

It is also possible to obtain $\epsilon^{2}$-expansion
of $ A(x,\epsilon,k) $ from the expansion (\ref{pervsyst}),
which corresponds to quasi-classical limit for
Schr\"odinger operator:

\begin{equation}
\label{razla}
A(x,\epsilon,k) = \sum_{n=0}^{\infty} \epsilon^{2n}
{\hat A}_{n} (x,k)
\end{equation}
where ${\hat A}_{0} (x,k) = {1 \over \sqrt{k^{2} + 2 u(x)} }$, and

$$ {\hat A}_{1} (x,k) = {1 \over 2 \sqrt{k^{2} + 2 u(x)} }
\left[{1 \over 4} ({\hat A}_{0x})^{2} - {1 \over 2}
{\hat A}_{0} {\hat A}_{0xx} \right] ,$$

$${\hat A}_{n} (x,k) =$$ 
$$= {1 \over 2 \sqrt{k^{2} + 2 u(x)} }
\left[{1 \over 4} \sum_{s=0}^{n-1} 
\left( {d \over dx} {\hat A}_{s} \right)
\left( {d \over dx} {\hat A}_{n-s-1} \right) -
{1 \over 2} \sum_{s=0}^{n-1} {\hat A}_{s}
{d^{2} \over dx^{2}} {\hat A}_{n-s-1} \right] - $$
\begin{equation}
\label{rec1}
- {1 \over 2} \sqrt{k^{2} + 2 u(x)}
\left[ \sum_{s=1}^{n-1} {\hat A}_{s} {\hat A}_{n-s} \right],
\,\,\, n \geq 2 .
\end{equation}

As can be easily seen, any ${\hat A}_{n}(x,k)$ is the expression
containing only polynomials of $u,u_{x},\dots$ divided by some odd
degrees of $\sqrt{k^{2} + 2 u(x)}$:

$$ {\hat A}_{n} (x,k) = \sum_{q=1}^{3n} 
{ {\hat D}_{n}^{q} (u,u_{x},\dots,u_{2nx}) \over
(\sqrt{k^{2} + 2 u(x)})^{2q+1} } \,\,\,\,\, , \,\,\, n \geq 1 . $$

Let us now consider the first formal equation of (\ref{pervsyst})
in the form of the formal expansion:

$$ {\partial u \over \partial \tau} = \lambda 
\sum_{n=1}^{\infty} \alpha_{n} \gamma_{n-1} \partial_{x}
{1 \over \pi i} \oint (2k)^{2n} \left[ \sum_{s=0}^{\infty}
\epsilon^{2s} \sum_{q=0}^{3s}
{ {\hat D}_{s}^{q} (u,u_{x},\dots,u_{2sx}) \over
(\sqrt{k^{2} + 2 u(x)})^{2q+1} }\right] dk $$
where ${\hat D}_{s}^{q} (u,u_{x},\dots,u_{2sx})$ do not depend
on k and formal integration

$$ {1 \over \pi i} \oint (2k)^{2n} 
{dk \over (\sqrt{k^{2} + 2 u(x)})^{2q+1} } $$
 coincides here with the value of this integral.

The value 

$$ \lambda \sum_{n=1}^{\infty} 
\alpha_{n} \gamma_{n-1} {1 \over \pi i} \oint (2k)^{2n} 
{dk \over \sqrt{k^{2} + 2 u(x)}} $$

coincides by the definition with $\int \Phi (u)du$ because
${\hat A}_{0}(x,k) = 1/ \sqrt{k^{2} + 2 u(x)}$
and the limit of every KdV-equation at 
$\epsilon \rightarrow 0$ is:

$$ {\partial u \over \partial \tau} = 
{u^{n-1} u_{x} \over \alpha_{n}} $$

so we must have ${\partial u \over \partial \tau} = \Phi (u) u_{x},\,\,\,
\epsilon = 0$, and any of the values

\begin{equation}
\label{qval}
\lambda \sum_{n=1}^{\infty} 
\alpha_{n} \gamma_{n-1} {1 \over \pi i} \oint (2k)^{2n}
{dk \over (\sqrt{k^{2} + 2 u(x)})^{2q+1} }
\end{equation}
is equal by such a way to

$$ {(-1)^{q} \over (2q-1)!!} {d^{q} \over du^{q} }
\int \Phi (u) du = {(-1)^{q} \over (2q-1)!!} \Phi^{(q-1)} (u(x)), \,\,\,
q \geq 1 . $$ 

Using these two equalities in the first equation of (\ref{pervsyst})
we obtain the equation (\ref{epsuravn}) in the required form in
any order of $\epsilon^{2}$ after the finite number of steps
(\ref{rec1}).

Theorem 3 is proved.

Corollary.

If $\Phi (u)$ is such that the equation

\begin{equation}
\label{nul}
x - u_{0}(x,\tau) + \tau \Phi (u_{0}(x,\tau)) \equiv 0
\end{equation}
has the unique solution for any $x$ and $0 \leq \tau \leq 1$
then $u(x,\tau,\epsilon)$ can be represented as the formal
expansion in powers of $\epsilon^{2}$:

\begin{equation}
\label{uexp}
u(x,\tau,\epsilon) = u_{0}(x,\tau) + \sum_{n=1}^{\infty}
\epsilon^{2n} u_{n}(x,\tau) , \,\,\, -\infty < x < \infty , \,\,\,
0 \leq \tau \leq 1 ,
\end{equation}
where $u_{0}(x,\tau)$ satisfies (\ref{nul}).

The proof is evident since we have a linear non-homogeneous
evolution equation
on every $u_{n}(x,\tau)$ with the initial data 
$u_{n}(x,0) \equiv 0$ which always has a unique solution.

\begin{center}
{\bf Recurrent formulas for $u_{n}(x,1)$ in the $\epsilon^{2}$ -
expansion of $u(x,1)$.}
\end{center}

Theorem 4.

Let $\Phi (u)$ be a convergent everywhere series:
$\Phi (u) = \sum_{n=0}^{\infty} \gamma_{n} u^{n}$
such that the equation (\ref{nul}) has a unique solution
for any $x$ and $0 \leq \tau \leq 1$, then the solution 
$u(x) \equiv u(x,1)$ of (\ref{utauur})

$$ {\partial u \over \partial \tau} = \sum_{n=1}^{\infty}
\alpha_{n} \gamma_{n-1} \partial_{x} {\delta \over \delta u(x)}
\int \chi_{2n+1}(x) dx $$
(formal Euler-Lagrange derivative) with the initial data:
$u(x,0) \equiv x$ can be represented as the formal expansion
in terms of $\epsilon^{2}$

\begin{equation}
\label{uuu}
u(x) = u_{0}(x) + \sum_{n=1}^{\infty} \epsilon^{2n}
u_{n}(x)
\end{equation}

where $u_{0}(x)$ satisfies

\begin{equation}
\label{nul1}
x - u_{0}(x) + \Phi (u_{0}(x)) =0
\end{equation}

and the coefficients $u_{n}(x)$ can be found from
the recurrent formulas:

$$A_{0}(x,k) = {1 \over  \sqrt{k^{2} + 2 u_{0}(x)}}$$

$$ u_{1}(x) = u_{0x} {\hat L} < 
{1 \over 2 \sqrt{k^{2} + 2 u_{0}(x)}} \left(
{1 \over 4} (A_{0x})^{2} - {1 \over 2} A_{0}
A_{0xx} \right) >$$

$$A_{1}(x,k) = {1 \over 2 \sqrt{k^{2} + 2 u_{0}(x)}} \left(  
{1 \over 4} (A_{0x})^{2} - {1 \over 2} A_{0}
A_{0xx} \right)$$

$$u_{n}(x) = u_{0x} {\hat L} <
{1 \over 2 \sqrt{k^{2} + 2 u_{0}(x)}} \{
{1 \over 4} \sum_{s=0}^{n-1} 
\left({d \over dx} A_{s}\right)
\left({d \over dx} A_{n-s-1}\right) -$$
$$- {1 \over 2} \sum_{s=0}^{n-1} A_{s}
{d^{2} \over dx^{2}} A_{n-s-1} -
(k^{2} + 2 u_{0}(x)) \sum_{s=1}^{n-1}
A_{s} A_{n-s} -$$ 
\begin{equation}
\label{uexp1}
- 2 \sum_{z=1}^{n-1} u_{z}(x)
\left( \sum_{s=0}^{n-z} A_{s} A_{n-z-s} \right)
\} > \,\,\,\,\,\,\,\,\, , n \geq 2
\end{equation}

$$A_{n}(x,k) = {1 \over 2 \sqrt{k^{2} + 2 u_{0}(x)}} \{
{1 \over 4} \sum_{s=0}^{n-1}
\left({d \over dx} A_{s}\right)
\left({d \over dx} A_{n-s-1}\right) -$$
$$ - {1 \over 2} \sum_{s=0}^{n-1} A_{s}
{d^{2} \over dx^{2}} A_{n-s-1} -
(k^{2} + 2 u_{0}(x)) \sum_{s=1}^{n-1}
A_{s} A_{n-s} - $$
\begin{equation}
\label{aexp1}
- 2 \sum_{z=1}^{n-1} u_{z}(x)
\left( \sum_{s=0}^{n-z} A_{s} A_{n-z-s} \right)
\} - { u_{n}(x) \over (\sqrt{k^{2} + 2 u_{0}(x)})^{3} }
\,\,\,\,\,\,\,\, , n \geq 2
\end{equation}
where all $A_{n}(x,k)$ have the form

\begin{equation} 
\label{aform}
A_{n}(x,k) = \sum_{q=1}^{3n} 
{D_{n}^{q} (u_{0},u_{0x},\dots,u_{0 (2n)x}) \over
(\sqrt{k^{2} + 2 u_{0}(x)})^{2q+1} }
\end{equation}

and ${\hat L}$ is the linear operator acting on the
functions of k so that:

\begin{equation}
\label{L0}
{\hat L} < {1 \over \sqrt{k^{2} + 2 u_{0}(x)} } >
= \Phi ( u_{0}(x) )
\end{equation}

\begin{equation}
\label{LL}
{\hat L} < {1 \over (\sqrt{k^{2} + 2 u_{0}(x)})^{2q+1} } >
= { (-1)^{q} \over (2q-1)!! } \Phi^{(q)} ( u_{0}(x) ) \,\,\, , q \geq 1 .
\end{equation}

(Let us note here that $A_{n}(x,k)$ can be obtained from the 
introduced previously $A_{s}^{*}(x,k)$ if we substitute
the function $u(x)$ in the form (\ref{uexp}) in all
the expressions for $A_{s}^{*}(x,k)$.)

Proof.

Let us change the first equation of (\ref{pervsyst})
by the equation $\Omega (x) \equiv 0$ at $\tau = 1$, where
$\Omega (x)$ is of the form (\ref{forma}) which is identically
zero at $t_{1} = t_{2} = \dots = 0$ ($u (x,0,0,\dots) = x$)

\begin{equation}
\label{novom}
\Omega (x) = x - u(x) + \sum_{s=1}^{\infty} \beta_{s} 
{\delta \over \delta u(x)} \int \chi_{2s-1}(x,\epsilon) dx .
\end{equation}

It is not very difficult to check using (\ref{obint}) that
for $\tau = 1$ the coefficients $\beta_{n}$, corresponding to the flux
(\ref{utauur}) can
be expressed in terms of introduced in
(\ref{kdvtype}) and (\ref{ffunction}) $\alpha_{n}$ and
$\gamma_{n}$ by formula

$$ \beta_{n} = - 4 (2n - 1) \alpha_{n} \gamma_{n-1} ,$$
and, as follows from (\ref{sootnint})
$\alpha_{n} = (-1)^{n}(n+1)! / 4^{n}n(2n-1)!!$

According to the formula (\ref{dubrrel}) we can
write this equation in the form:

\begin{equation}
\label{wcw}
x - u(x,\epsilon) + \sum_{n=0}^{\infty} \lambda \beta_{n} 
{1 \over \pi i} \oint (2k)^{2n-2} A(x,k,\epsilon) dk =0 
\end{equation}
where $\beta_{n}$ are such that:

\begin{equation}
\label{sum}
\sum_{n=1}^{\infty} \lambda \beta_{n}                 
{1 \over \pi i} \oint (2k)^{2n-2} { dk \over 
\sqrt{k^{2} + 2 u_{0}(x)} } \equiv
\Phi ( u_{0}(x) )
\end{equation}
for any $u_{0}(x)$ so that at $\epsilon = 0$ we obtain
formula (\ref{nul1}).

As can be easily shown using the formal representation (\ref{wcw})
(just like as in Theorem 3) $\Omega (x)$ 
can be represented as a formal
series on $\epsilon^{2}$, which at any power of $\epsilon^{2}$ is
just a local expression of $u,u_{x},u_{xx},\dots$, having the form:

\begin{equation}
\label{ddd}
\Omega (x) = \sum_{s=0}^{\infty} \epsilon^{2s} 
N(u,u_{x},\dots,u_{(2s)x})
\end{equation}
where $N(u,u_{x},\dots,u_{(2s)x})$ are polynomials of
$u_{x},u_{xx},\dots$ containing $2s$ derivatives with respect 
to $x$, with the coefficients depending on 
$\Phi (u), \Phi^{(1)}(u),\dots$ . This means that the formal
series (\ref{wcw}) in any order of $\epsilon^{2}$ is
convergent everywhere (as the sum of differentiations of convergent
everywhere series (\ref{sum}) ) and can be expressed in the 
appropriate form (\ref{ddd}).

It is evident from this fact that at any finite order of
$\epsilon^{2}$ $\Omega (x)$ satisfies to linear differential
equation of finite order like (\ref{omegaev}) according to
n-th KdV equation, since it is so for any finite sum (\ref{novom}),
and from Theorem 3 we can conclude that 
it is also so for the evolution
with respect to $\tau$.

So that, we can use the local equality $\Omega (x) \equiv 0$,
where $\Omega (x)$ is local expression of $u,u_{x},\dots$,
in any order of $\epsilon^{2}$ for $u(x)$ in the form (\ref{uuu})
at $\tau = 1$ if $u(x,\tau)$ is a formal global asymptotic
solution of (\ref{utauur}) for $0 \leq \tau \leq 1$.

Now let us introduce linear operator ${\hat L}$ acting on functions
of $k$ by the formula:

$${\hat L} < G(k) > = \sum_{n=1}^{\infty}
\lambda \beta_{n} {1 \over \pi i} \oint (2k)^{2n-2}
G(k) dk  $$

By the definition:

$$ {\hat L} < {1 \over \sqrt{k^{2} + 2 u_{0}(x)} } >
\equiv \Phi ( u_{0}(x) ) $$
and it can be easily seen that
$$ {\hat L} < 
{1 \over (\sqrt{k^{2} + 2 u_{0}(x)})^{2q+1} } >
= {(-1)^{q} \over (2q-1)!! } \Phi^{(q)} ( u_{0}(x) ) $$
where $\Phi^{(q)}(u) \equiv {d^{q} \over du^{q} } \Phi (u)$,
for $q \geq 1$.

By the substitution of expansions 

$$ A(x,k,\epsilon) = {1 \over \sqrt{k^{2} + 2 u_{0}(x)} }
+ \sum_{n=1}^{\infty} \epsilon^{2n} A_{n} (x,k) $$
and
$$ u(x,\epsilon) = u_{0}(x) + \sum_{n=1}^{\infty} \epsilon^{2n}
u_{n}(x) $$
in the system

$${1 \over 2} \epsilon^{2} A {d^{2} \over dx^{2} } A -
{1 \over 4} \epsilon^{2} \left( {d \over dx }A \right)^{2} =
1 - (k^{2} + 2 u(x)) A^{2} ,$$
$$ x - u(x,\epsilon) + {\hat L} < A(x,k,\epsilon) > 
= 0 $$
(that is $u_{n}(x) = {\hat L} < A_{n}(x,k) >$ ),
it is easy to obtain (\ref{uexp1}) and (\ref{aexp1}) for
$u_{n}$ and $A_{n}$, where we used the fact that

$$ {\hat L} <
{1 \over (\sqrt{k^{2} + 2 u_{0}(x)})^{3} } >
= - \Phi^{(1)} ( u_{0}(x) )  $$
and that view (\ref{nul1}) 
$\Phi^{(1)} ( u_{0}(x) ) = 1 - {1 \over u_{0x}}$.

Formulas (\ref{aform}) for $A_{n}$ are evident from the recurrent
formulas (\ref{aexp1}).

Theorem 4 is proved.

\vspace{1cm} 

As can be easily seen from (\ref{nul1}), all  
$\Phi^{(q)} ( u_{0}(x) )$
can be expressed in terms of $u_{0}(x)$ and its derivatives using
formula $\Phi^{(q+1)} ( u_{0}(x) ) = {1 \over u_{0x} } {d \over dx}
\Phi^{(q)} ( u_{0}(x) )$ and so it can be easily seen that we 
can represent $u_{n}(x)$ in the form:

\begin{equation}
\label{undr}
u_{n}(x) = \sum_{s=1}^{N(n)} 
{U_{n}^{s}(u_{0xx},u_{0xxx},\dots) \over (u_{0x})^{s} }
\end{equation}
where $U_{n}^{s}(u_{0xx},u_{0xxx},\dots)$ are polynomials
containing $(2n+s)$ differentiations with respect to $x$ in
each term.

All the functions $u_{n}(x) , n \geq 1$ 
(see ~\cite{EYY1}) are full double derivatives with respect to
$x$ of the functions $F_{n}(x)$, where 
$F_{1}(x) = (1/24)\ln (u_{0x})$ and all $F_{n}(x), n \geq 2$ have 
the same form as (\ref{undr}):

\begin{equation}
\label{fndr}
F_{n}(x) = \sum_{s=1}^{N(n)-2}
{f_{n}^{s}(u_{0xx},u_{0xxx},\dots) \over (u_{0x})^{s} } 
\end{equation}
The condition (\ref{fndr}) fixes here uniquely the integration 
constants in the determination of $F_{n}(x)$, such that

$$u_{n}(x) = {d^2 \over dx^2} F_{n}(x)$$

The values $F_{n}(x)$ in the form (\ref{fndr}) are necessary
for the calculations of Weil-Petersson volumes of moduli spaces
and it is quite easy to propose an algorithm for determination
of $F_{n}(x)$ using $u_{n}(x)$ in the form (\ref{undr}). Namely
we define the algorithm of integration of the expression
(\ref{undr}) provided that it is the full derivative with respect
to $x$ in the following way:

since $D^{-1} u_{n}$ must have the same form

\begin{equation}
\label{dndr}
D^{-1} u_{n} = \sum_{s=1}^{N(n)-1}
{v_{n}^{s}(u_{0xx},u_{0xxx},\dots) \over (u_{0x})^{s} } 
\end{equation}
where $v_{n}^{s}$ are polynomials (containing $(2n+s-1)$ 
differentiations with respect to $x$ in each term), it is
easy to define the highest term 
$v_{n}^{N(n)-1}(u_{0xx},u_{0xxx},\dots)$ from (\ref{undr})
as 

$$v_{n}^{N(n)-1} = {U_{n}^{N(n)}(u_{0xx},u_{0xxx},\dots)
\over (N(n) - 1) u_{0xx} } $$
which must be polynomial of $u_{0xx},u_{0xxx},\dots$ if
$u_{n}(x)$ is the full derivative of (\ref{dndr}).

The rest of (\ref{dndr}) without the highest term
$$\sum_{s=1}^{N(n)-2}
{v_{n}^{s}(u_{0xx},u_{0xxx},\dots) \over (u_{0x})^{s} }$$
is the integral of the value

$$u_{n}(x) - {U_{n}^{N(n)} \over (u_{0x})^{N(n)} }
- {( v_{n}^{N(n)-1} )_{x} \over (u_{0x})^{N(n)-1} } =
\sum_{s=1}^{N(n)-1} {U_{n}^{s}(u_{0xx},u_{0xxx},\dots) 
\over (u_{0x})^{s} } - {( v_{n}^{N(n)-1} )_{x} \over 
(u_{0x})^{N(n)-1} } $$
and we can repeat the same procedure for 
$ v_{n}^{N(n)-2}(u_{0xx},u_{0xxx},\dots)$ and so on.
The last step must give identically zero for the
corresponding rest of sum (\ref{undr}) for $u_{n}(x)$ -
full derivative of the expression (\ref{dndr}) and so we
shall obtain quite simple formulas for $v_{n}^{s}$ in terms
of $U_{n}^{m}$ and their derivative. Applying this procedure
twice to any $u_{n}(x)$ in the form (\ref{undr}) we shall
obtain the required form (\ref{fndr}) for the functions
$F_{n}(x)$.

\vspace{1cm}
The second author (A.Ya.M.) thanks INTAS 
(grant INTAS 96-0770) and RFBR (grant 97-01-00281) for
partial financial support.

\end{document}